\documentclass[preprint]{aastex7}

\newcommand{\Rs}{$ R_{\odot}$}
\usepackage{multirow}
\accepted{September 15, 2025}
\journalinfo{}
\graphicspath{{./}{figures/}}
\begin{document}

\title{Periodic Variations in Visible Light Brightness  as Tracers of Fine Coronal Structures}

\correspondingauthor{Nathalia Alzate}

\author[orcid=0000-0001-5207-9628,gname=Nathalia,sname=Alzate]{Nathalia Alzate}
\affiliation{NASA Goddard Space Flight Center, Greenbelt, MD, USA}
\affiliation{ADNET Systems, Inc., Greenbelt, MD, USA}
\email[show]{nathalia.alzate@nasa.gov}

\author[orcid=0000-0001-6407-7574,gname=Simone,sname=Di Matteo]{Simone Di Matteo}
\affiliation{Physics Department, The Catholic University of America, Washington, DC, USA}
\affiliation{NASA Goddard Space Flight Center, Greenbelt, MD, USA}
\email{simone.dimatteo@nasa.gov}

\begin{abstract}

The quiescent or dynamic nature of fine scale ray-like features in the sun corona, observed in visible light, is still an open question. Here, we show that \added{most of} daily and hourly periodic variations in visible light brightness of the high corona (up to 15 \Rs) are aligned to the tip of streamers and are consistent with the periodicity of plasma release from simulations of tearing-induced magnetic reconnection at the heliospheric current sheet. The areas in which we detect periodicities can be used as tracers of non-quiescent fine coronal rays. This also allows their distinction from coronal rays more likely to be real quiescent features or associated with smaller and/or faster unresolved brightness variations. In the low- and middle-corona (down to 1.4 \Rs) similar brightness variations are observed \added{along loop-like and cusp-like features marking boundaries of streamers which then connect to radial features in the high corona. This suggests the presence of additional mechanisms in the low- and middle-corona periodically releasing density structures in the solar wind}. The periodicity \added{distributions show a} solar cycle \added{modulation with shorter periods (smaller structures) during solar maximum}. \added{Periodicities} are observed \added{within} streamers during solar minimum, but are visible at all latitudes, even extending radially from the poles, during solar maximum.

% \textbf{Articles published in ApJ Letters must also be concise and to the point, and most should fall within the following limits:}
% \begin{itemize}
%     \item Abstract – no more than 250 words
%     \item Main Text – no more than 3500 words (not including acknowledgments, appendices or other supplementary material)
%     \item Figures and Tables – no more than 5 combined figures (each limited to 9 panels) and tables, e.g. 3 figures and 2 tables.
% \end{itemize}

% \textbf{However, these are no longer intended to be compulsory, and new manuscripts that exceed these limits can be considered at the discretion of the Scientific Editor; in exceptional circumstances, such as results of very broad significance, ApJ Letters will publish substantially longer articles. In any case, sufficient introductory background material should be included, and the content of the article should be generally understandable by scientists who are not specialists in the particular field. If in doubt, authors are encouraged to contact the ApJ Letters Editor prior to submission.}

\end{abstract}

% \keywords{\uat{Galaxies}{573} --- \uat{Cosmology}{343} --- \uat{High Energy astrophysics}{739} --- \uat{Interstellar medium}{847} --- \uat{Stellar astronomy}{1583} --- \uat{Solar physics}{1476}}



%%%%%%%%%%%%%%%%%%%%%%%%%%%%%%%%%%%%%%%%%%%%%%%%%%%%%%%%%%%%%%%%%
\section{Introduction}
\label{sec:intro}
%%%%%%%%%%%%%%%%%%%%%%%%%%%%%%%%%%%%%%%%%%%%%%%%%%%%%%%%%%%%%%%%%

Coronal rays appear in visible light as bright narrow lanes corresponding to electron density enhancements, superimposed within, or composing larger-scale dense streamers. These structures have been clearly observed during total solar eclipses and, more routinely, from space-based coronagraphs and imagers. Coronal rays extend for several tens of solar radii (\Rs), appear as quiescent regions with respect to large dynamic transient features in the solar corona (e.g., coronal mass ejections or CMEs), and are typically associated with the interplanetary sector boundary \citep{Hansen1974, Howard1974, Wang1997, Wang2000, Liewer2023, Saez2005, Thernisien2006}. These properties suggest a relation between the coronal rays and the quiet regions in which the slow solar wind forms. Coronal rays show a significant variability on yearly timescales mostly due to solar cycle reconfiguration of the coronal magnetic field topology \citep{Golub2009}. Some variability occurs down to daily time scales while only small variations \added{have been reported} at hourly timescales \citep{Guhathakurta1995}. One possible interpretation is that coronal rays are quasi-static structures such that the daily changes are mostly related to the solar rotation bringing the rays in the plane of sky (POS) of the observing telescope. In this scenario, single rays could be due to folds/wrinkles in the heliospheric current sheet (HCS) \citep{Liewer2023} or to inhomogeneous distribution of open magnetic flux tubes \citep{Ko2022}. These scenarios were invoked by \citet{Poirier2020} to explain the fine structures and coronal rays resolved in observations from the Wide-Field Imager (WISPR; \citealt{Howard2019}) onboard Parker Solar Probe (PSP; \citealt{Raouafi2014}). Another explanation could be that resolving underlying small-scale and continuous dynamics in current instruments is challenging. Therefore, results based on the assumption of quiescent structures on sub-streamer size scales and hourly (and smaller) time scales can be misleading. 

In fact, streamers show considerable temporal variation of density, and are host to several small-scale dynamic features \citep{DeForest2018, Sheeley1997, Viall2015} that have been observed to propagate from the corona into the solar wind. In particular, helmet streamer plasmoids (``Sheeley blobs'') are structures that place constraints on the acceleration and source of the slow solar wind. There is strong evidence that this type of transient originates above 2 \Rs\ through the pinching-off of expanding helmet-streamer loops, sometimes in a periodic manner \citep{Viall2015, Sanchez-Diaz2017, Reville2020}. On smaller (sub-streamer) scales, the presence of transients has been suggested to be due to the occurrence of short period (down to a few minutes) periodic density structures from in situ density and composition observations at 1 AU \citep{Kepko2020, Gershkovich2023,Kepko2024}. Additionally, sub-streamer features have been reported to propagate from the low corona below 2 \Rs, in overlapping extreme ultraviolet-white light (EUV-WL) observations, to the outer corona in WL \citep{Alzate2023, Alzate2024}. Remote sensing and in situ data suggest that much of the structures observed in the slow solar wind are a tracer of solar wind formation (e.g., \citealt{Alzate2017, Viall2008, Viall2009, Viall2010, Viall2015, Sheeley1997, Sheeley2007, Sheeley2009, Rouillard2009, Rouillard2010, Rouillard2011, Arge2010, DiMatteo2019}). Close to the solar surface, jetlets \citep{Raouafi2023, Raouafi2016, Raouafi2014} have been suggested as the source of ``microstreams" and were found to correspond to structured solar wind observed in situ by PSP \citep{Kumar2023, Kumar2022}. Evidence of the structured solar wind is further supported by observations of transients in WISPR \citep{Poirier2023, Reville2020} and METIS \citep{Ventura2023} remote sensing observations. The interpretation of these transients spans a variety of phenomena: small-scale “blobs”, signatures of wave propagation and switchbacks \citep{Efimov2012,Miyamoto2014,Shoda2021,Chiba2022,Andretta2025}, all of which might be related to driving from below (e.g., supergranules in the photosphere), magnetic reconnection at null points, or interchange reconnection between open/closed magnetic field lines \citep{Drake2021, Morgan2013, Poirier2023}.

In this paper, we \added{investigate the relationship between coronal rays and quasi-periodic plasma outflows. We} present \added{advanced image processing techniques to reveal quiescent and dynamic fine structures in the solar corona as well as comprehensive} observations of periodicities \added{from hourly to daily timescales} in visible light observations from 1.4 to 15 \Rs\ \added{throughout Solar Cycle 24}.  In Section \ref{sec:datameth} we describe the specifics of the observations and of the methods used to carry out this analysis. Section \ref{sec:observations} presents our results followed by the discussion and our conclusion in Sections \ref{sec:discussion} and \ref{sec:conclusions}.

%%%%%%%%%%%%%%%%%%%%%%%%%%%%%%%%%%%%%%%%%%%%%%%%%%%%%%%%%%%%%%%%%
\section{Data and Methods} 
\label{sec:datameth}
%%%%%%%%%%%%%%%%%%%%%%%%%%%%%%%%%%%%%%%%%%%%%%%%%%%%%%%%%%%%%%%%%

\added{Here, we provide an overview of the datasets and techniques we use to reveal quiescent and dynamic fine structures in the solar corona. Particularly, our targets are fine scale radial structures (coronal rays) present in white light observations of the solar corona. We implement a spectral analysis approach to test for the occurrence of periodic brightness variations at each pixel possibly related to the transit of quasi-periodic plasma outflows, which have been invoked as possible building blocks or inherent features of coronal rays. Our analysis is complemented by advanced image processing techniques to reveal these fine scale structures as well as the large scale features of the solar atmosphere (e.g., streamers) in which rays occur.}

%%%%%%%%%%%
\subsection{Datasets}
\label{sec:datasets}
\begin{figure}[t]
    \centering
    \includegraphics[width=0.9\linewidth]{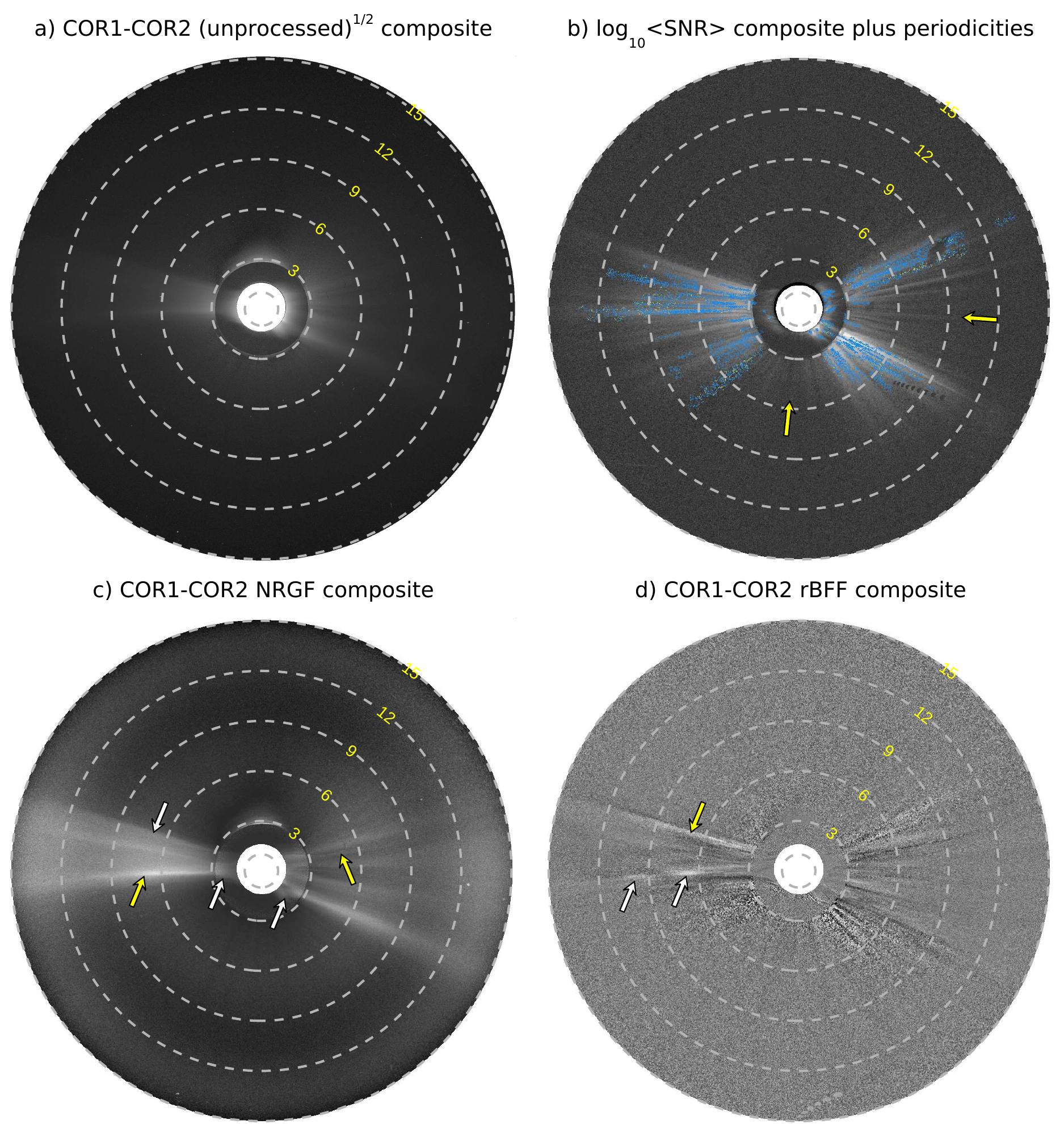}
    \caption{\footnotesize\textbf{Time series spectral analysis and image processing techniques reveal fine scale structures in coronagraph observations.} a) Square root brightness of unprocessed COR1-COR2 composite for 17 January at 13:52 UT; b) periodicities detected for the six-day period of 16-22 January 2008 (blue and orange dots) over the $log_{10}<SNR>$ summary image (gray scale) representing $f=9.6-36.5 \mu Hz$ frequency range corresponding to 7.6-28.9 hr; c) NRGF-processed COR1-COR2 composite for 17 January at 13:52 UT; d) rBFF-processed COR1-COR2 composite for 17 January at 13:52 UT.}
    \label{fig:image_proc_examples}
\end{figure}

%%%%%%%%%%%
We used data from the Sun Earth Connection and Heliospheric Investigations \citep[SECCHI,][]{Howard2008} suite of instruments on board the Solar Terrestrial Relations Observatory Ahead \citep[STEREO-A;][]{Kaiser2008} twin spacecraft. Specifically, we used total polarized brightness observations by the COR1 inner coronagraph \citep{Thompson2003}, and the COR2 outer coronagraph \citep{Howard2008}. The COR1 instrument observes the corona in a field of view (FOV) \added{between} 1.4 and 4.0 $R_{\odot}$ and COR2 observes between 2.5 and 15 $R_{\odot}$. Here and throughout this paper, we refer to heliocentric heights in solar radii \added{as the Cartesian distance units in the plane-of-sky of coronagraphs FOV. An example of a composite of the square root of unprocessed COR1-COR2 observations is in Figure \ref{fig:image_proc_examples}}. In this study, we use 1024x1024 \added{pixel COR2 and 512x512 pixel COR1} images for \added{five time intervals of $T=$6 days and cadence, $\Delta t$, listed in Table \ref{tab:results}}. \added{The time intervals and cadence were chosen based on the data availability of COR2 ensuring less than 5\% of missing files and no more than two consecutive missing images. Then, we decimated the COR1 observations to have the same cadence. For the later years (last event in Table \ref{tab:results}), the COR1 observations is significantly affected by the degradation of the instrument \citep{Thompson2018}. To mitigate this effect, we stacked 12 images at 5 min cadence, resulting in a frame per hour.}

\begin{table}[t]
   \footnotesize
   \centering
    \begin{tabular}{|c|c|c|cccccc|}
        Time interval & $\Delta t$ & Solar Cycle & \multicolumn{6}{c|}{Frequency Bands [$\mu$Hz]}\\
         & [min] &  & \multicolumn{6}{c|}{(Period Bands [hr])}\\
    \hline
      2008/01/16 19:50 - 2008/01/22 19:50 & 30 & minimum &  15.4$_{9.6}^{21.1}$ & 26.0$_{21.1}^{30.8}$ & 36.5$_{30.8}^{42.3}$ & 52.8$_{42.3}^{63.4}$ & 82.7$_{63.4}^{101.9}$ & 184.6$_{101.9}^{267.2}$\\
       & & & (18.1$_{13.1}^{28.9}$ & 10.7$_{9.0}^{13.1}$ & 7.6$_{6.6}^{9.0}$ & 5.3$_{4.4}^{6.6}$ & 3.4$_{2.7}^{4.4}$ & 1.5$_{1.0}^{2.7}$) \\
      2008/12/11 00:20 - 2008/12/17 00:20 & 30 & minimum & 15.4$_{9.6}^{21.1}$ & 
      26.0$_{21.1}^{30.8}$ & 36.5$_{30.8}^{42.3}$ & 52.8$_{42.3}^{63.4}$ & 82.7$_{63.4}^{101.9}$ & 184.6$_{101.9}^{267.2}$ \\
       & & & (18.1$_{13.1}^{28.9}$ & 10.7$_{9.0}^{13.1}$ & 7.6$_{6.6}^{9.0}$ & 5.3$_{4.4}^{6.6}$ & 3.4$_{2.7}^{4.4}$ & 1.5$_{1.0}^{2.7}$) \\
      2011/02/03 00:05 - 2011/02/09 00:05 & 60 & ascending & 11.6$_{9.6}^{13.5}$ & 18.3$_{13.5}^{23.1}$ & 31.8$_{23.1}^{40.4}$ & 51.0$_{40.4}^{61.5}$ & 78.8$_{61.5}^{96.1}$ & 112.2$_{96.1}^{128.4}$ \\
       & & & (24.1$_{20.6}^{28.9}$ & 15.2$_{12.0}^{20.6}$ & 8.7$_{6.9}^{12.0}$ & 5.5$_{4.5}^{6.9}$ & 3.5$_{2.9}^{4.5}$ & 2.5$_{2.2}^{2.9}$) \\
      2012/08/11 00:05 - 2012/08/17 00:05 & 60 & maximum & 11.6$_{9.6}^{13.5}$ & 19.2$_{13.5}^{25.0}$ & 31.7$_{25.0}^{38.4}$ & 48.1$_{38.4}^{57.7}$ & 71.2$_{57.7}^{84.6}$ & 106.5$_{84.6}^{128.4}$ \\
       & & & (24.1$_{20.6}^{28.9}$ & 14.4$_{11.1}^{20.6}$ & 8.8$_{7.2}^{11.1}$ & 5.8$_{4.8}^{7.2}$ & 3.9$_{3.3}^{4.8}$ & 2.6$_{2.2}^{3.3}$) \\
      2016/06/27 00:05 - 2016/07/03 00:05 & 60$^a$ & descending & 11.6$_{9.6}^{13.5}$ & 17.3$_{13.5}^{21.1}$ & 29.8$_{21.1}^{38.4}$ & 48.1$_{38.4}^{57.7}$ & 70.2$_{57.7}^{82.7}$ & 105.5$_{82.7}^{128.4}$ \\
       & & & (24.1$_{20.6}^{28.9}$ & 16.1$_{13.1}^{20.6}$ & 9.3$_{7.2}^{13.1}$ & 5.8$_{4.8}^{7.2}$ & 4.0$_{3.4}^{4.8}$ & 2.6$_{2.2}^{3.4}$) \\
    \end{tabular}
    \caption{\footnotesize The five 6-day intervals analyzed: time range, selected cadence of the instruments, phase of the solar cycle, identified preferred frequency bands of periodic brightness variations (center values with subscripts and superscripts indicating the band limits). $^a$ The COR1 images were stacked for this event.}
    \label{tab:results}
\end{table}

%%%%%%%%%%%
\subsection{Time Series Spectral Analysis}
\label{sec:spectral}
%%%%%%%%%%%
\added{Some coronal rays have been associated with the} continuous release of transients \citep{DeForest2018,Poirier2023}\added{, sometimes in a periodic manner. This property determines brightness variations along the coronal ray which could be captured by analyzing the time series spectral properties at each pixel of the image. Therefore, we investigated the possibility of using the power spectrum and/or the occurrence of periodic variations as a proxy for the identification of fine coronal structures}. We perform a spectral analysis of the brightness time series resulting from each pixel independently. \added{We use an updated version of the spectral analysis procedure used by \citet{Viall2015} to estimate the power spectral density (PSD), via the adaptive multitaper method \citep[aMTM;][]{Thomson1982} but refined by combining additional approaches \citep{Mann1996,Vaughan2005} which make the technique suitable for the analysis of astrophysical and geophysical time series \citep[\textit{SPD\_MTM;}][]{DiMatteo2021}.} \added{More detailed information on the spectral analysis procedure is in Appendix \ref{sec:appendix_spec}. For this work, we considered time intervals of $T=$6 days with a cadence of either $\Delta t=$30 or $\Delta t=$60 min, which correspond to Rayleigh frequency of $f_{Ray}=1/T=1/(N\Delta t)=1.9 \mu Hz$ and Nyquist frequency ($f_{Ny}=1/(2\Delta t)$) of either $\approx139\mu Hz$ or $\approx278 \mu Hz$. For the spectral analysis we considered time-halfbandwidth product $NW=2.5$, $K=4$ tapers, and a 90\% confidence threshold for the amplitude test and amplitude+F test.

The spectrum of coronagraph time series is characterized by a red spectrum at low frequency which progressively flattens due to shot noise \citep{Telloni2013,Threlfall2017}. Particularly, \citet{Telloni2013} showed that the low frequency power, not necessarily associated with periodic variations \citep{Threlfall2017}, can reveal fine structures in the solar corona (e.g., coronal rays). Here, we identify regions of enhanced power by evaluating an average Signal-to-Noise Ratio as:

\begin{equation}
    <SNR> = \frac{\left (\sum_{{i=s}}^{{s+S}} PSD(f_i) \cdot \Delta f \right)/S\Delta f}{\left (\sum_{j=m}^{m+M} PSD_{bkg}(f_j) \cdot \Delta f \right)/M\Delta f}=\frac{\left (\sum_{i=s}^{s+S} PSD(f_i) \right)/S}{\left (\sum_{j=m}^{m+M} PSD_{bkg}(f_j)\right)/M}
\end{equation}

where we perform a simple rectangular integration to estimate the power for the \emph{signal}, in the frequency band $[f_{s},f_{s+S}]$ of the estimated PSD, and \emph{noise}, in the frequency band $[f_{m},f_{m+M}]$ of the identified PSD background ($PSD_{bkg}$). Both frequency bands are within the ``safe'' frequency range (see Appendix \ref{sec:appendix_spec}) and the step in Fourier frequencies, $f_i$, is $\Delta f$ and equal to the Rayleigh frequency, $f_{Ray}$. An example of a logarithmic $<SNR>$ image is in Figure \ref{fig:image_proc_examples}b in which we consider as \emph{signal} the variations on time scales between 28.9 and 6.9 hr and \emph{noise} below 2.5 hr. The $<SNR>$ image provides a summary of the areas spanned by bright features over the course of the time interval under consideration and reveals fine and faint radial structures (i.e., coronal rays) not only within streamers but also at high latitude regions as marked by yellow arrows.  

Finally, our spectral analysis approach can further identify the location of pixels associated with quasi-periodic brightness variations according to two independent tests. In Figure \ref{fig:image_proc_examples}b we indicate the pixels at which we identified periodic variations according to the amplitude test (blue) and amplitude+F test (orange) in the  $f=$ 9.6--40.4 $\mu Hz$ frequency range, corresponding to $\approx$ 6.9--28.9 hr (see Appendix \ref{sec:appendix_refine} for a full description of the steps of our procedure). The locations of these pixels cluster along fine quasi-radial structures in coronagraph observations providing an additional way to identify possible coronal rays. Additionally, the comparison with the $<SNR>$ image enables the distinction between rays with or without quasi-periodic brightness variations.
}

%%%%%%%%%%%
\subsection{Image Processing}
\label{sec:imgproc}
%%%%%%%%%%%
For each event, we investigate the general context of the corona revealed by the Normalizing Radial Graded Filter (NRGF) method \citep{Morgan2006}. \added{The NRGF is a multiplicative filter that, after evaluating the brightness average and standard deviation of sun-centered annulus in the image, uses their radial profiles, estimated through interpolation, to standardize the brightness values. Figure \ref{fig:image_proc_examples} shows a comparison between an unprocessed (panel a) and a NRGF processed (panel c) image. By} removing the steep radial gradient\added{, the NRGF filter reveals mainly the large scale} electron corona structures\added{, like streamers (white arrows), but can also highlight radial structures, mainly the streamers' stalk and finer structures, possibly coronal rays (yellow arrows)}. The spatio-temporal variations are instead highlighted by processing the images with the Bandpass Frame Filtering (BFF) method \citep{Alzate2021,Alzate2025}. The \added{BFF is a temporal bandpass filter applied to a datacube formed by a suite of frames uniformly distributed in time, treating brightness values from each pixel as a singular time series. The} temporal bandpass filter effectively damps high-frequency noise and low-frequency slow-changing structures. The filtering is achieved through convolution with two normalized \added{Gaussian} kernels defined as a wide \added{and narrow Gaussian kernel based on the corresponding standard deviation, $\sigma$, defining the limit of the frequency range of interest. The choice is based on the definition of cut-off frequency, $f_c$, for a Gaussian filter leading to $\sigma = 1/(2\pi\Delta tf_c)$ where we chose $T_c=1/f_c=$30 hr for the wide filter and $T_c=$2.5 hr for the narrow filter. Here, we adopt a modified version of the BFF filter. First, we divided, instead of subtracted, the original observations by the wide filtered image which compensates the brightness drop with distance. The narrow filter is applied as in the original BFF procedure. The resultant ratio-BFF image \citep[$I_{rBFF, }$][]{Alzate2025}, has brightness values around one, therefore, we can enhance features in the processed images via a hyperbolic tangent function centered around one, namely, $I^{'}_{rBFF} = tanh(8(I_{rBFF}-1))$. We also take advantage of the $<SNR>$ values to enhance the contrast between areas with high and low $<SNR>$, namely, $I^{''}_{rBFF} = I^{'}_{rBFF}(1+2.5\,tanh(<SNR>-1))$. An example of such rBFF processed images is in panel d, in which we show examples of plasma parcels moving out from the sun (white arrows) and fine scale radial structures (yellow arrow) likely related to coronal rays. Note how they appear within the large scale features revealed by the NRGF processed images.} For this study, the NRGF and \added{r}BFF methods are used in tandem to characterize fine scale structures of the corona, particularly coronal rays, which might be associated with quiescent regions as well as with continuous release of transients \citep{DeForest2018,Poirier2023}.

%%%%%%%%%%%%%%%%%%%%%%%%%%%%%%%%%%%%%%%%%%%%%%%%%%%%%%%%%%%%%%%%%
\section{Observations} 
\label{sec:observations}
%%%%%%%%%%%%%%%%%%%%%%%%%%%%%%%%%%%%%%%%%%%%%%%%%%%%%%%%%%%%%%%%%
\added{The first step in our analysis is to identify periodic brightness variations in visible light. Particularly, we are interested in the possible occurrence at preferential frequency bands and their location, from the low- to the high-corona. Once characteristic time scales and locations are identified, the comparison with processed COR1 and COR2 observations can provide more information on the relation between quiescent and dynamic features.}

\begin{figure}
    \centering
    \includegraphics[width=0.5\linewidth]{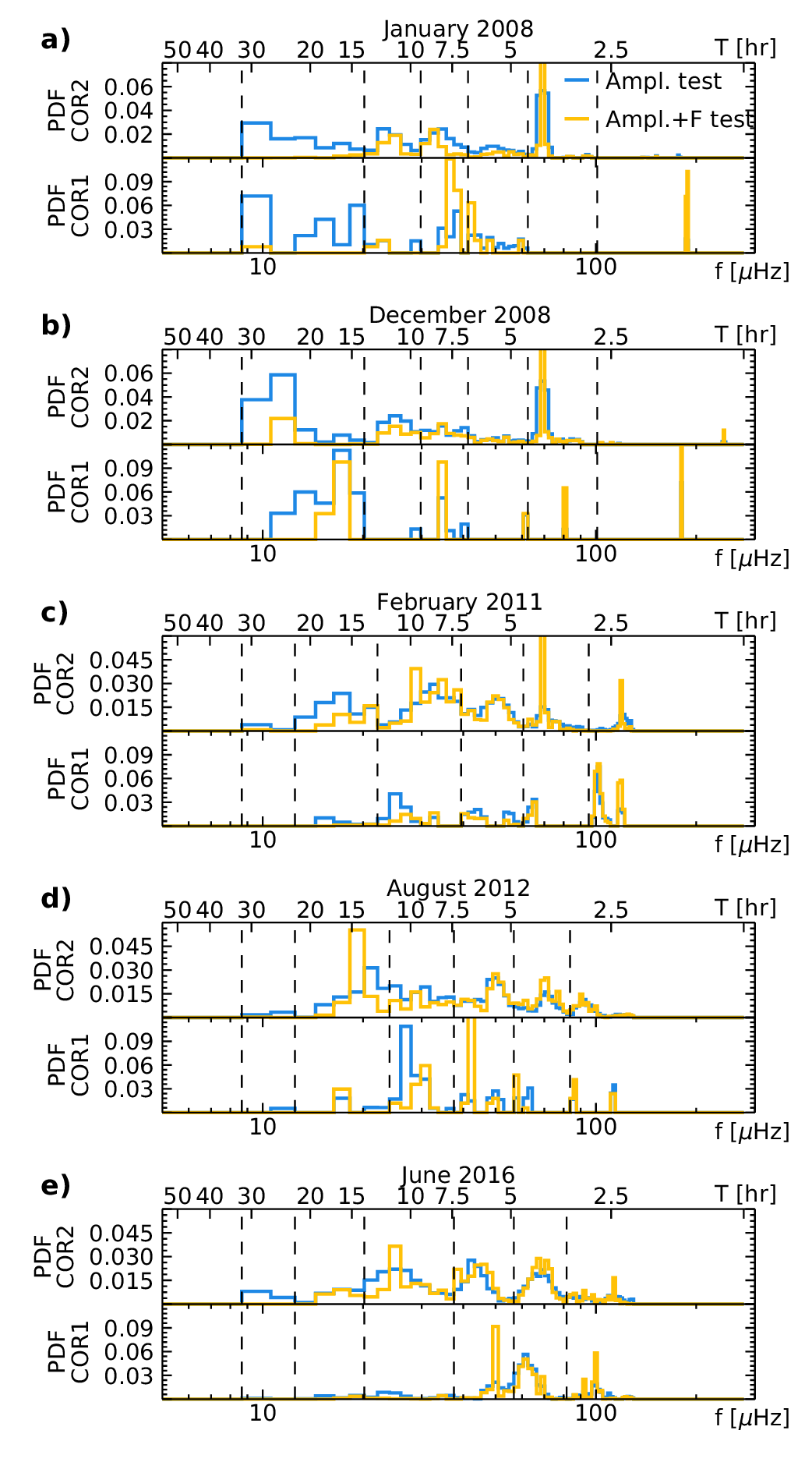}
    \caption{\footnotesize\textbf{The periodicity distributions reveal preferential bands and a modulation with the solar cycle.} For the five events under investigation, panels a--e show the PDF of the number of pixels at which periodicities were detected according to the amplitude test (blue lines) and amplitude+F test (orange lines) in COR2 (top panel) and COR1 (bottom panel) observations.}
    \label{fig:PDF_freq}
\end{figure}

\subsection{Frequency (period) distribution of quasi-periodic visible light brightness variations}
\added{Previous observations and simulations estimated the release of quasi-periodic density structures at $\approx$90--180 min, $\approx$7--10 hr, and $\approx$25--50 hr timescales \citep[e.g.,][]{Poirier2023}. Our choice of spectral analysis parameters allows the investigation of periodicity in a ``safe'' frequency range (see Appendix \ref{sec:appendix_spec}) from 9.6 $\mu$Hz to 128.4 $\mu$Hz (from $\approx$2.2 hr to $\approx$29.0 hr). Note that for our first two events the upper frequency limit is higher reaching 267.2 $\mu$Hz ($\approx$1.0 hr). In Figure \ref{fig:PDF_freq} we show the probability density function (PDF) of the number of pixels at which a periodicity is detected. For each event, the blue profiles indicate the results from the amplitude test and the orange ones from the amplitude+F test. In each panel, the results for COR2 are in the top row and COR1 in the bottom row. The PDFs show the presence of preferential frequency bands, delimited by dashed vertical lines whose values are reported in Table \ref{tab:results}. In the results from COR2 observations, we can recognize preferential frequency bands at $\approx$20--30 hr, $\approx$7--9 hr, and below $\approx$3 hr (consistent with previous observations and simulations) for all events, but more distinct during the solar minimum ones (panel a and b). Noteworthy is that for the solar minimum events, during which we can investigate higher frequencies, we also detected $\approx$90 min periodicities. However, our results show the existence of a complementary spectrum of quasi-periodic brightness variations at $\approx$10--20 hr, $\approx$5--7 hr, and $\approx$3--5 hr more evident during the solar maximum event (panel d). In the results for COR1 observations, we observe preferential frequency bands at $\approx$13--30 hr and $\approx$7--9 hr with less evident complementary bands during solar minimum events (panel a and b). For the other events, periodicities are mostly observed between 5 and 15 hr.}

\subsection{Coronal rays and fine structures from $<SNR>$ images}
\begin{figure}
    \centering
    % \begin{interactive}{animation}{Fig3_movie.mp4}
    \includegraphics[width=\linewidth]{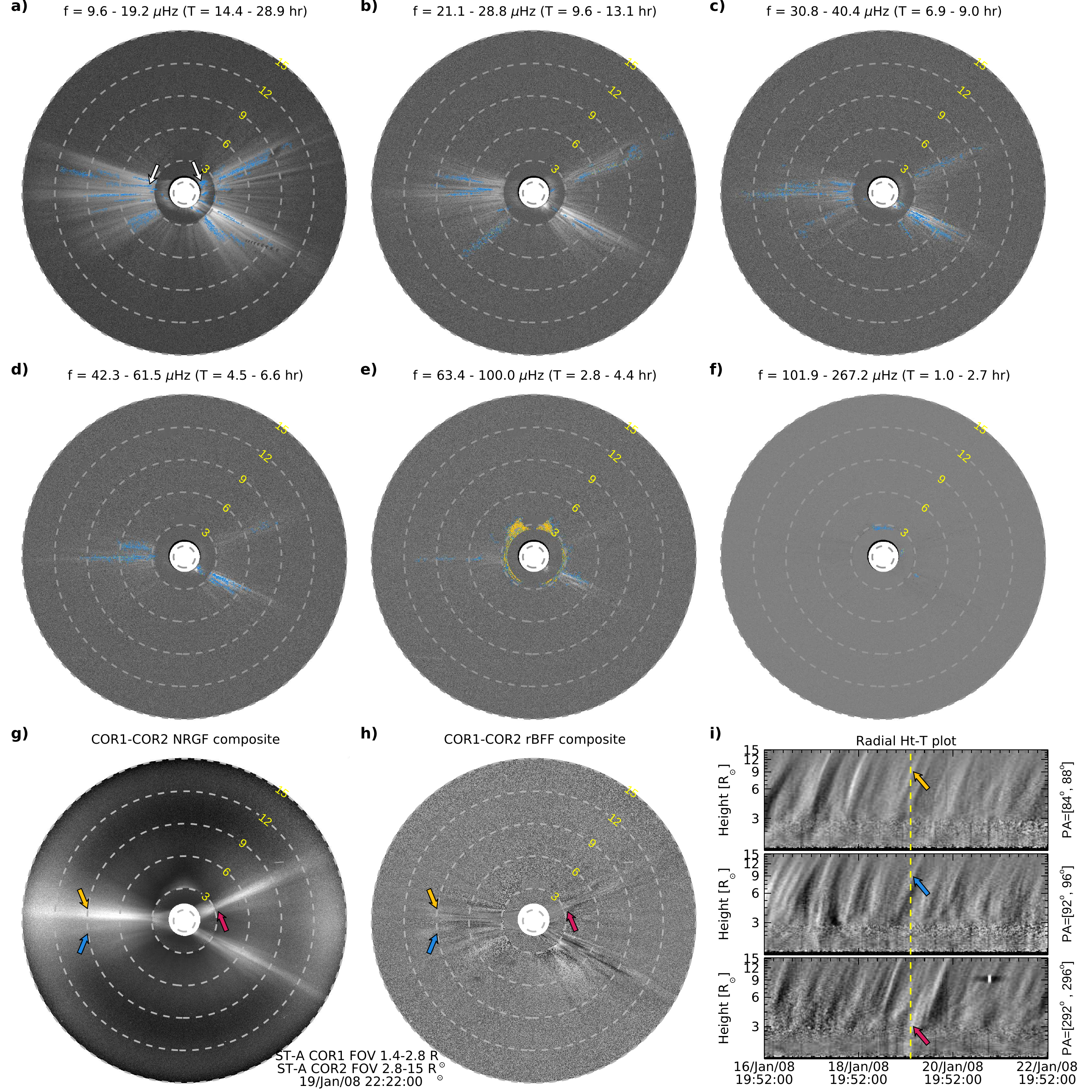}
    % \end{interactive}
    \caption{\footnotesize\textbf{Fine scale coronal structures revealed by time series spectral analysis procedures and image processing techniques for the January 2008 event (solar minimum).} Panels a--f show the $<SNR>$ images (gray scale) and the periodic brightness variations detected via the amplitude test (blue dots) and amplitude+F test (orange dots) revealing fine scale ray-like, cusp-like, and loop-like features (examples indicated by white arrows) within the frequency bands identified in Figure \ref{fig:PDF_freq}. Example of periodically released structures (marked by arrows) on 2008 January 19 at 22:22 UT in a COR1-COR2 composite processed with the NRGF method (panel g) and rBFF method (panel h). Panel i shows a six day Ht-T plot of average brightness along a 4$^\circ$-wide radial slice including the path of structures identified in panel h, for which the corresponding time is marked by the yellow dashed line. The associated animation of NRGF and rBFF processed COR1-COR2 composites with 30-min cadence spans 2008 January 16–22 (8 s total duration).}
    \label{fig:summary_2008Jan}
\end{figure}

\begin{figure}
    \centering
    % \begin{interactive}{animation}{Fig4_movie.mp4}
    \includegraphics[width=\linewidth]{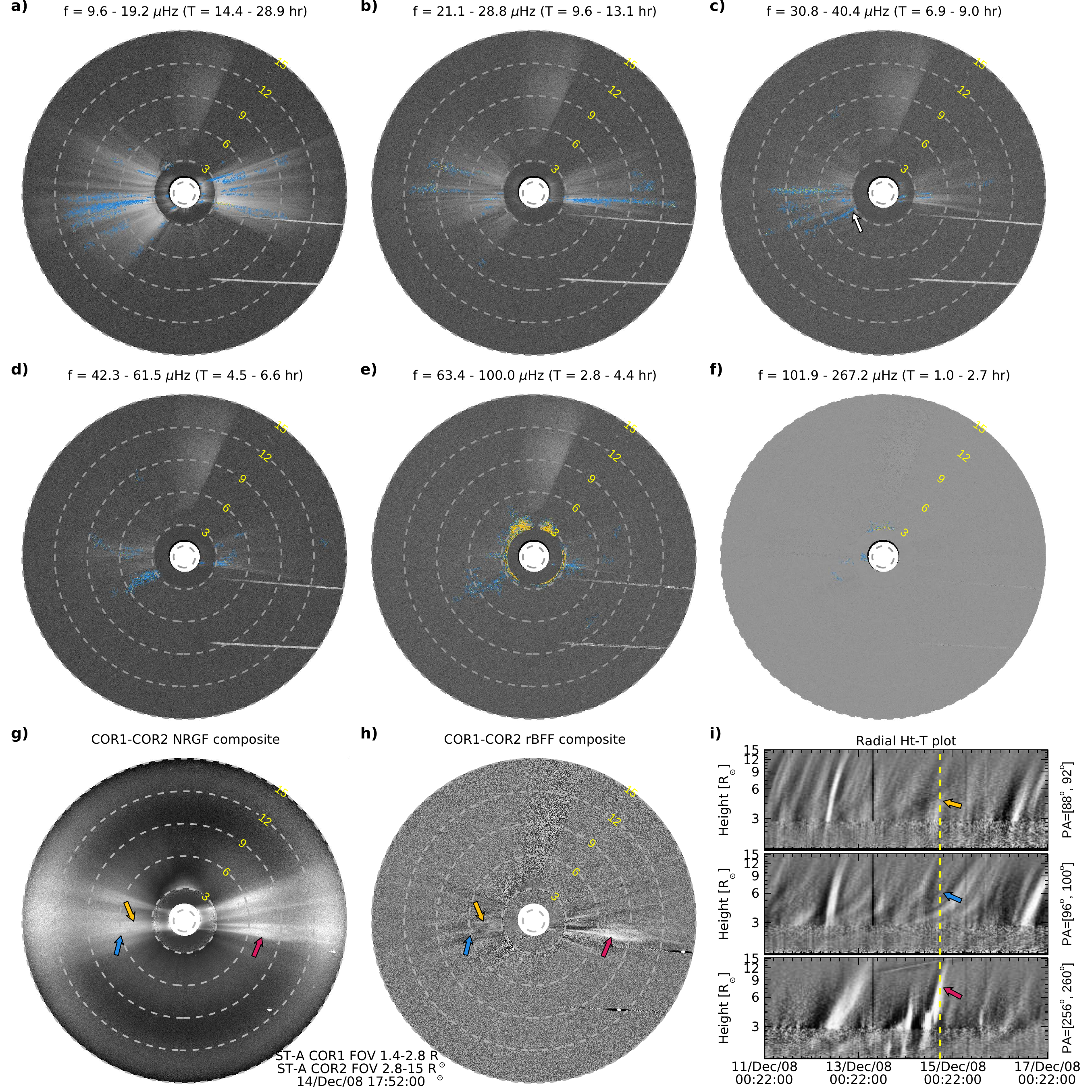}
    % \end{interactive}
    \caption{\footnotesize Same display of Figure \ref{fig:summary_2008Jan} but for the 2008 December event (solar minimum). The associated animation of NRGF and rBFF processed COR1-COR2 composites with 30-min cadence spans 2008 December 11–17 (8 s total duration).}
    \label{fig:summary_2008Dec}
\end{figure}

\begin{figure}
    \centering
    % \begin{interactive}{animation}{Fig5_movie.mp4}
    \includegraphics[width=\linewidth]{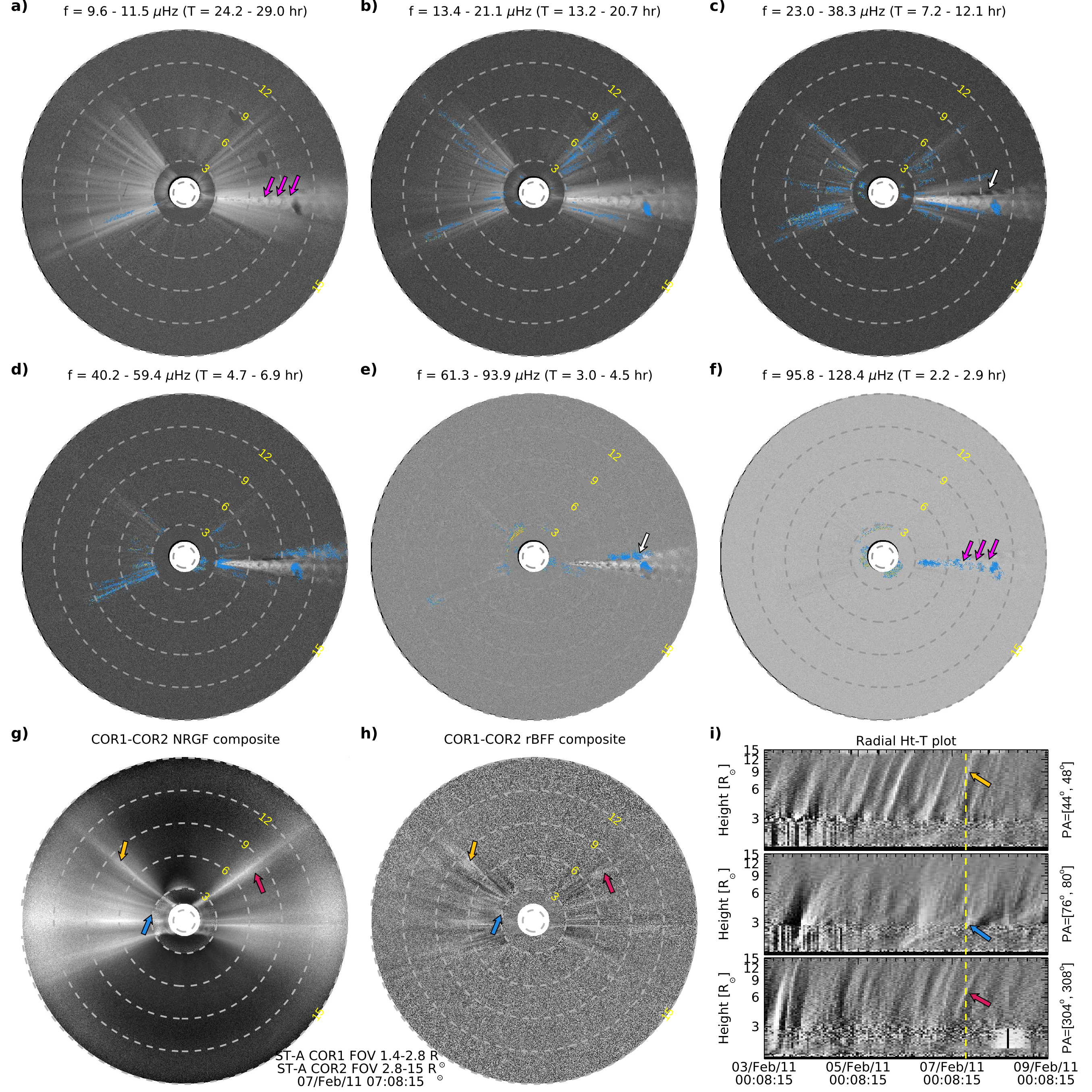}
    % \end{interactive}
    \caption{\footnotesize Same display of Figure \ref{fig:summary_2008Jan} but for the 2011 February event (ascending phase of the solar cycle). The associated animation of NRGF and rBFF processed COR1-COR2 composites with 60-min cadence spans 2011 February 3–9 (8 s total duration).}
    \label{fig:summary_2011Feb}
\end{figure}

\begin{figure}
    \centering
    % \begin{interactive}{animation}{Fig6_movie.mp4}
    \includegraphics[width=\linewidth]{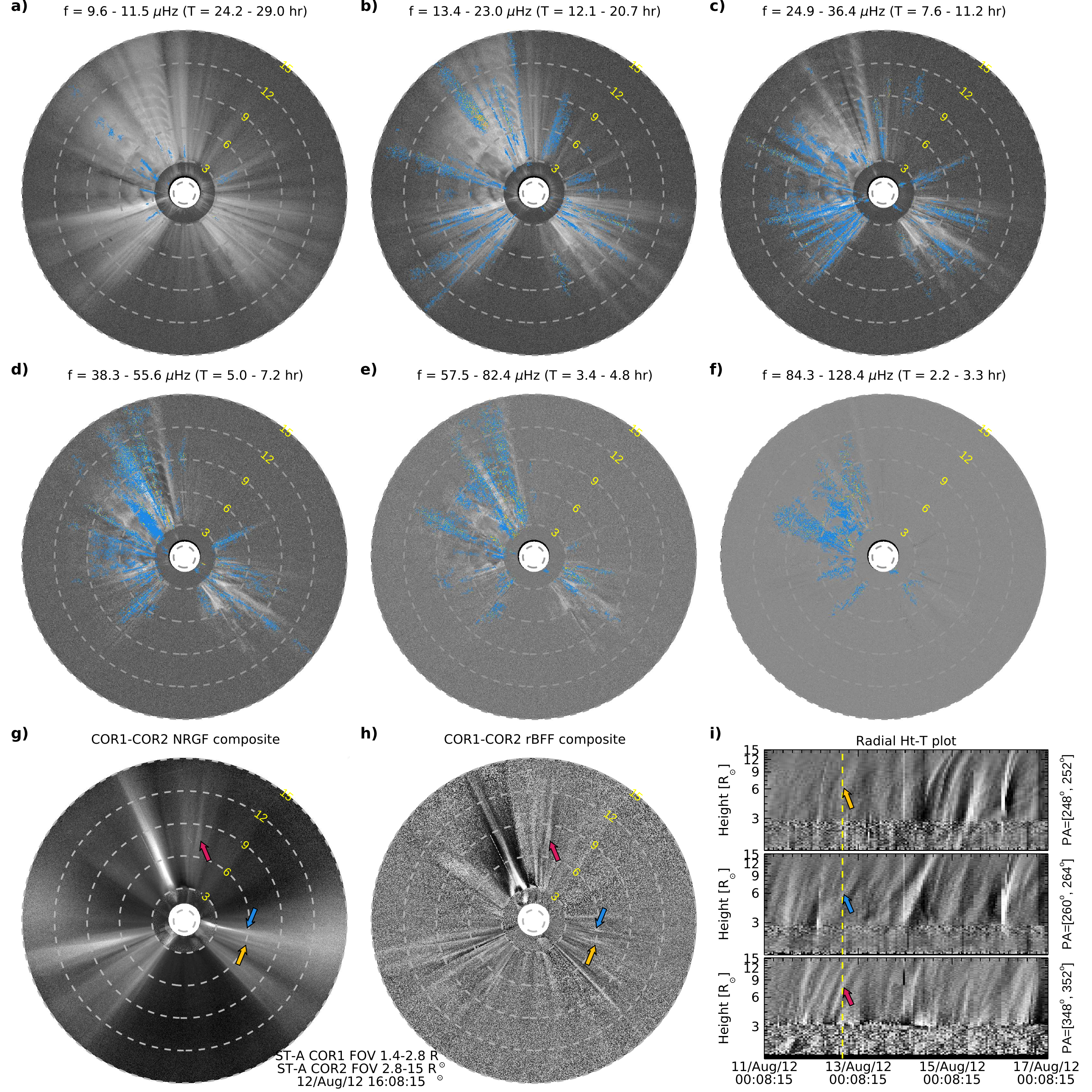}
    % \end{interactive}
    \caption{\footnotesize Same display of Figure \ref{fig:summary_2008Jan} but for the 2012 August event (solar maximum). The associated animation of NRGF and rBFF processed COR1-COR2 composites with 60-min cadence spans 2012 August 11–17 (8 s total duration).}
    \label{fig:summary_2012Aug}
\end{figure}

\begin{figure}
    \centering
    % \begin{interactive}{animation}{Fig7_movie.mp4}
    \includegraphics[width=\linewidth]{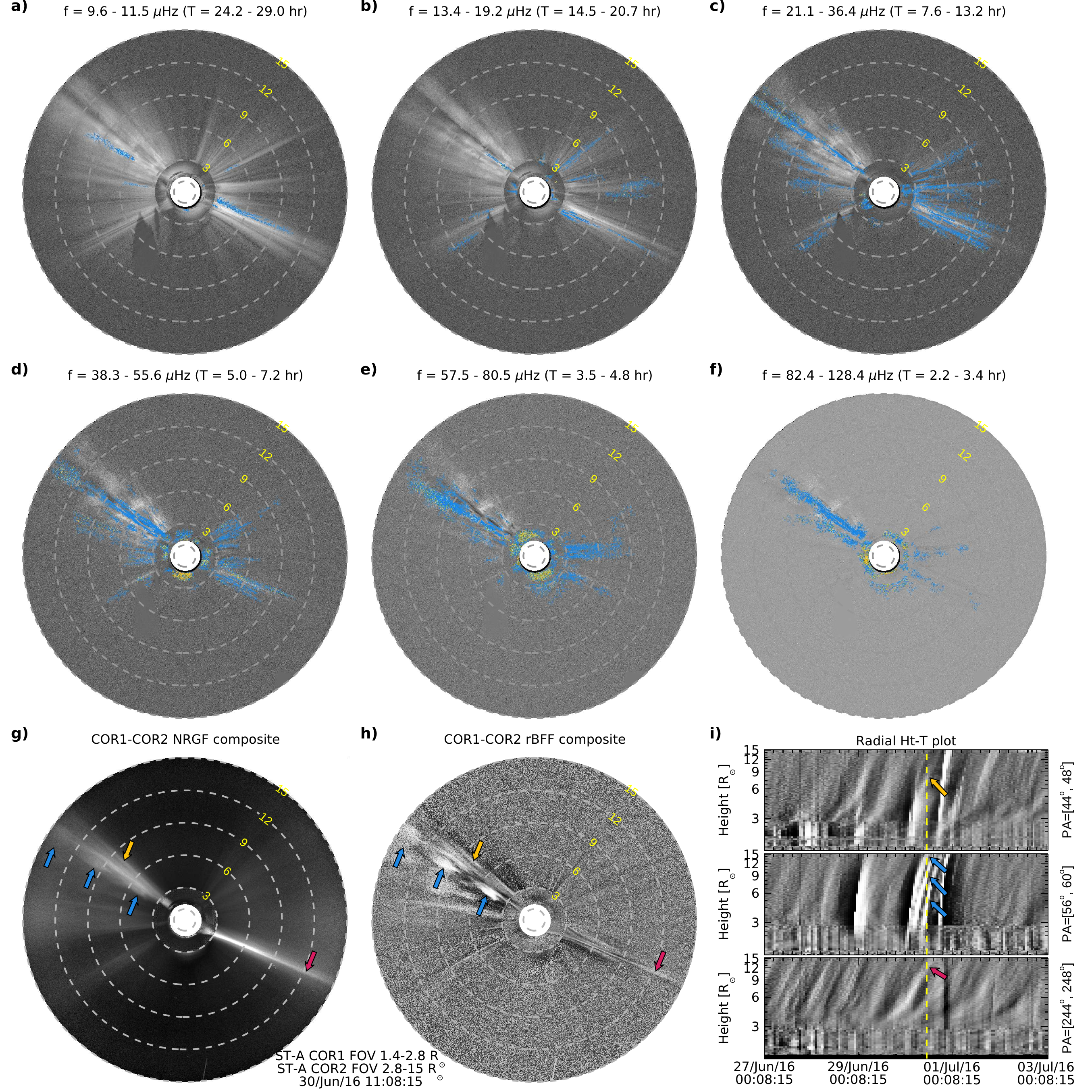}
    % \end{interactive}
    \caption{\footnotesize Same display of Figure \ref{fig:summary_2008Jan} but for the 2016 June event (descending phase of the solar cycle). The associated animation of NRGF and rBFF processed COR1-COR2 composites with 60-min cadence spans 2016 June 27 – July 3 (8 s total duration).}
    \label{fig:summary_2016Jun}
\end{figure}

\added{Before analyzing the location of the pixels at which periodicities were identified, we investigated the distribution of the $<SNR>$ which we show are related to fine coronal structures. For the January 2008 event, panels a--f in Figure \ref{fig:summary_2008Jan} show the COR1-COR2 composite of logarithmic $<SNR>$ images for frequency bands reported at the top of each panel. For time scales between 14.4 and 28.9 hr (panel a), the bright areas (high $<SNR>$ values) indicate the locations spanned by large scale structures: below 3 \Rs\ we can recognize the cusp-like features corresponding to streamers; above 3 \Rs\ we observe both wide and very narrow radial features related to streamer stalks and coronal rays. At lower time scales (panel b--e), the shape of large scale features is progressively lost in exchange for more detailed visualization of very narrow radial features, mostly above 3 \Rs, with additional loop-like feature closer to the inner edge of COR1. Finally, when reaching the range dominated by shot noise (panel f), the $<SNR>$ image appears mostly uniform. The same characteristics appear in the results for the December 2008 event, panels a--f in Figure \ref{fig:summary_2008Dec}, occurring during solar minimum like the previous event. For the 2011 February event in Figure \ref{fig:summary_2011Feb}, during the ascending phase of the solar cycle, we observed similar results but with some fine scale features appearing also in panel f for the lowest timescales. For this event, we have an isolated CME erupting from the west limb (close to the equator) which appears as a sudden increase in the brightness time series of pixels crossed by the ejecta. Sudden jumps in a time series constitute a broad band source of power that increases the power spectrum values at all frequencies. As a consequence all the $<SNR>$ images show the CME profile as it crosses the COR1-COR2 FOV (i.e., areas marked by magenta arrows), progressively highlighting substructures of the CME moving from low to high frequency. While panel a shows a more uniform CME front, panel f shows brighter portions of the CME (this is confirmed by the associated animation). For the 2012 August event in Figure \ref{fig:summary_2012Aug}, during solar maximum, radial features comprise the entire FOV covering also the polar region, with cusp-like and loop-like features clearly outlined below $\approx$3\Rs\ (best seen in panel a). During this 6-day period there were multiple CME eruptions with their outlines visible in all the $<SNR>$ images, though overlapped since occurring in similar spatial regions (the various CMEs are visible in the associated animation). For the 2016 June event in Figure \ref{fig:summary_2016Jun}, during the declining phase, the fine structures revealed in the $<SNR>$ images are closer to the equator, with the signature of the passage of a CME in the east limb.}

\subsection{Coronal rays and fine structures from periodicities in visible light brightness}

\added{We then investigated the location of the pixels at which we identified periodic brightness variations overlaid on the $<SNR>$ images in Figures \ref{fig:summary_2008Jan}--\ref{fig:summary_2016Jun}. The blue dots indicate the pixels at which the amplitude test is satisfied at a frequency within the range reported at the top of each panel; the orange dots instead refer to the results from the amplitude+F test (see Appendix \ref{sec:appendix_refine} for more details). For the two events during solar minimum (Figures \ref{fig:summary_2008Jan} and \ref{fig:summary_2008Dec}), the low frequency periodicities (period greater than 6.9 hr; panels a--c) are observed from the low- to the middle- and high-corona. Below $\approx$3 \Rs, the periodicities occur either along non-radial boundaries of cusp-like features, often connecting to radial features above $\approx$3 \Rs, or loop-like features within them. Above $\approx$3 \Rs, the periodicities lay along radial fine-scale structures also revealed in the $<SNR>$ images. There are only few instances in which the periodicities at these heights lay on non-radial features, for example in the areas marked by white arrows in Figure \ref{fig:summary_2008Jan}a and Figure \ref{fig:summary_2008Dec}c. Collectively, in the COR1 FOV, identification of periodicities are sparse but always corresponding to portions of a streamer. In the COR2 FOV, periodicities are identified along almost all fine features revealed in the $<SNR>$ images except for the ones at the poles. Along faint features between the brighter regions, detection of periodicities rarely extend to the outer corona but are limited to heights of a few \Rs. Additionally, near the inner edge of the COR2 FOV, periodicity at $\approx$4 hours is detected at pixels forming concentric circles which is related to an instrumental signal already reported in previous investigations \citep{Viall2015,Alzate2024}.

The results for the event during the ascending phase of the solar cycle are similar (Figure \ref{fig:summary_2011Feb}). Some noteworthy elements are broad irregular patches (mostly from the blue dots of the amplitude test) better observed above the west limb in panels b--f along the path of a CME. We note that these patches match specular dark areas in the $<SNR>$ images at lower frequency (see for example blue patches in panel e and dark patches in panel c marked by white arrows). A more detailed analysis (not shown) of the corresponding brightness time series reveals that the periodicities identified in these regions are an artifact stemming from sudden brightness increase due to the passage of the CME front and internal structures or bright artifacts in the images. However, the features along the CME path resembling loops or cusps in COR1 and rays in COR2 are associated to real periodic variations.

These observations become relevant to interpret the result of the August 2012 event (solar maximum) during which there were multiple CMEs visible in the animation associated with Figure \ref{fig:summary_2012Aug}. Broad irregular patches, mostly observed at high frequencies (panels d--f) and along the CMEs' path, are likely false positives. However, we note a significant number of radial continuous features within these regions, also clearly identified by the amplitude+F test (orange), which are associated with real periodic brightness variations. These radial features, particularly relevant at lower frequencies (panels a--c), are identified at all latitudes except near the south pole where radial features show low $<SNR>$ values (panels a and b).

Finally, for the June 2016 event (Figure \ref{fig:summary_2016Jun}), during the descending phase of the solar cycle, the results are similar to the ones during the ascending phase (Figure \ref{fig:summary_2011Feb}). Note the loop-like and cusp-like features in the COR1 FOV extending to radial features in the COR2 FOV (panel b and c). As in previous events, we note broad patches, especially at high frequencies, along the path of a CME. At high frequencies (panels d--f) the results in the COR1 FOV appear distributed at all latitudes with only occasional correspondence to features in the COR2 FOV. However, we note that the more recent COR1 observations are affected by the instrument degradation \citep{Thompson2018}; in fact, we needed to stack images over an interval of 60 minutes for this analysis. Either instrumental limitation or the stacking might have affected the results of the spectral analysis results for COR1 in panels d--f.
}

\subsection{Comparison with processed COR1 and COR2 observations}
\added{We compare the location in which periodic brightness variations occur with the quiescent features revealed by the NRGF method and the fine scale dynamics revealed by the rBFF method (panels g--h from Figure \ref{fig:summary_2008Jan} to Figure \ref{fig:summary_2016Jun}). The arrows point at blob-like features in the rBFF images mapping to streamer stalks or faint radial features (coronal rays) in the NRGF images. We also generate height-time (Ht-T) plots averaging the brightness in rBFF images along radial slices (4$^\circ$ latitudinal width) centered on the indicated blob-like features. The Ht-T plots show that these regions, in which we detect periodicities, are associated with a continuous outflow of structures. Interestingly, such correspondence extends to polar latitudes during solar maximum as shown in Figure \ref{fig:summary_2012Aug}h where the red arrow marks a blob stemming from a polar coronal ray. This suggests that during active periods either polar rays also show periodicities or that the HCS is highly inclined and/or reaches polar latitudes. Alternative processes that can generate periodicities were also observed even though less frequently. During the same event in Figure \ref{fig:summary_2012Aug}h, a CME is erupting lower in the corona. The associated animation shows that the streamer stalk, after the passage of the CME, manifests clear wave-like activity (streamer waves) which typically occur on time scales of hours \citep[e.g.,][]{Decraemer2020}. Other examples are the yellow and blue arrows in Figure \ref{fig:summary_2016Jun}i where the structures are likely post eruption blobs along the CME path.}

%%%%%%%%%%%%%%%%%%%%%%%%%%%%%%%%%%%%%%%%%%%%%%%%%%%%%%%%%%%%%%%%%
\section{Discussion} 
\label{sec:discussion}
%%%%%%%%%%%%%%%%%%%%%%%%%%%%%%%%%%%%%%%%%%%%%%%%%%%%%%%%%%%%%%%%%
\added{We provide a comprehensive analysis of long (from hours to tens of hours) periodic visible light brightness variations from the low- and middle-corona to the high-corona for five events spanning a solar cycle. We provide additional evidence that original reports of periodic fluctuations might have been a characterization of the continuous power spectrum of non-periodic brightness variations due to an inadequate assumption of the background spectrum \citep{Telloni2013,Threlfall2017}. With our spectral analysis approach, we are able to distinguish regions associated with actual periodicities from the ones characterized by non-periodic background processes with power higher than the one of shot noise. 

The frequency distribution of the identified events reveal overall peaks limited to within preferred period bands of $\approx$20--30 hr, $\approx$10--20 hr, $\approx$7--9 hr, $\approx$5-7 hr, $\approx$3--5 hr, and more sporadically below $\approx$3 hr. These results include the periodicities expected from simulations of tearing-induced magnetic reconnection at the HCS, 25–50 h, 7–10 h, and 90–180 min \citep[e.g.,][]{Poirier2023}, but also reveal the presence of complementary bands. The period distributions for the various events also show a possible solar cycle modulation. During solar minimum, the periodicities show higher concentrations above $\approx$20 hr and at $\approx$7.5--15 hr; when approaching the solar maximum the distribution shifts toward shorter periodicities. This is consistent with previous reports of periodic variations detected in remote-sensing observations showing a similar trend from solar minimum and ascending phase \citep[e.g.,][]{Viall2015,Sanchez-Diaz2017,Ventura2023,Alzate2024} to solar maximum \citep[e.g.,][]{DeForest2018}.

Analysis of the location at which these periodicities are identified reveals that long periodicities (more than a few hours) tend to define fine structures (loop-like, cusp-like, and radial) often related to streamers. The short periodicities are distributed along wider and more irregular regions which are often associated with some form of ejecta, at least for our small sample of events (animations spanning the 6 days of each event are associated with Figures \ref{fig:summary_2008Dec}--\ref{fig:summary_2016Jun} and show CMEs during solar maximum and small streamer blowouts during solar minimum). These events disrupt the topology of the coronal magnetic fields, which can lead to oscillations of the streamer stalk \citep{Chen2010,Feng2011,Decraemer2020,Alzate2023}, determine velocity shears that can trigger instabilities \citep{Ofman_2011,Foullon_2011,Paouris_2024}, or lead to post-eruption blobs and plasmoids from reconnection in the current sheet trailing CMEs \citep[e.g., ][]{Lin2005}.

While most of these structures appear to form at $\approx$3 Rs, we also have examples in which the brightness track in the Ht-T plot can be traced down to $\approx$2 Rs (red arrow track in Figure \ref{fig:summary_2008Jan}, yellow and red arrow track in Figure \ref{fig:summary_2008Dec}). Note also that the formation at a height of $\approx$3 Rs, at least for some cases, can be an artifact due to the non-radial motion of features in the low- and middle-corona \citep{Alzate2024} which can lead a feature to be in or out of the radial slice during its propagation. This is also supported by the location of periodicities detected in the COR1 FOV which delineate non-radial features delimiting the boundary of streamer-like structures which converge toward radial features higher in the corona.} This is consistent with the results of \citealt{Alzate2024}, who identified brightness propagating disturbances originating more often at the boundary and within streamers in the low- and middle-corona. The clear shift of the areas of periodicities to the tip of streamers in the middle- and high-corona further supports that \added{most of} these long period coronal plasma releases are related to the tearing-induced magnetic reconnection at the HCS. Moreover, \added{some of} the observed periodicities match closely the ones expected from simulations (90--180 min, 7--10 h, and 25--50 h from \citealt{Poirier2023}). However, observations of periodicities along the streamer boundary in the low-corona suggest that \added{additional processes (e.g., }interchange reconnection\added{)} at the open and closed magnetic field \added{boundary} might also have \added{a} role in the release of these structures.

%%%%%%%%%%%%%%%%%%%%%%%%%%%%%%%%%%%%%%%%%%%%%%%%%%%%%%%%%%%%%%%%%
\section{Conclusions} 
\label{sec:conclusions}
%%%%%%%%%%%%%%%%%%%%%%%%%%%%%%%%%%%%%%%%%%%%%%%%%%%%%%%%%%%%%%%%%
Analysis of visible light observations from 1.4 to 15 \Rs, in the STEREO/COR1-COR2 FOV, reveals the presence of periodic brightness variations \added{in all phases of the Solar Cycle 24. We observe periodicities at $\approx$20--30 hr, $\approx$7--9 hr, and below $\approx$3 hr consistent with simulations of periodic formation of structures from tearing-induced magnetic reconnection at the HCS. Accordingly, most of the pixels at which a periodicity is identified are arranged along radial features (coronal rays) stemming from the tip of streamers.

However, our analysis reveals the presence of complementary periodicity bands at $\approx$10--20 hr, $\approx$5--7, and $\approx$3--5 hr. Additionally, areas characterized by periodicities extend to the low-corona along cusp-like features marking the boundaries of streamers. These results suggest the presence of additional mechanisms periodically releasing structures into the solar wind with interchange reconnection at the boundary of streamers being a likely candidate.}

During solar minimum, coronal rays are visible at all latitudes\added{, but only the ones with high $<SNR>$ values, often connected to the streamers, show periodic brightness variations.} During solar maximum, periodicities are observed up to the sun poles suggesting that during active periods either polar rays also show periodicities or that the HCS is highly inclined and/or reaches polar latitudes. \added{Additionally, our results confirm the presence of a solar cycle modulation of the observed periodicities with shorter period (smaller structures) during maximum solar activity.}

The outflows of quasi-periodic density structures at daily and hourly timescales \added{are supported in most of the} coronal rays in relation to streamers. The remaining coronal rays could still be related to unresolved outflows of structures or brightness variations either non-periodic or at shorter periodicities that we are not able to resolve. Alternatively, these coronal rays can  be quiescent structures with variations due to folds or inhomogeneous plasma distributions. More detailed analysis on higher resolution images and longer time periods is currently underway to better answer these questions.

\begin{acknowledgments}
N.A. acknowledges support from NASA ROSES through HGI grant No. 80NSSC20K1070, PSP-GI grant No. 80NSSC21K1945 and HSR grant No. 80NSSC25K7746. S.D. acknowledges support under the NASA ROSES PSP-GI grant No. 80NSSC21K1945 and HSR grant No. 80NSSC25K7746. The authors thank H. Morgan for valuable discussions about this work. \added{We thank the anonymous reviewer for the insight in identifying the valuable results of this study and helping to improve the quality of the work.}
\end{acknowledgments}

%%%%%%%%%%%%%%%%%%%%%%%%%%%%%%%%%%%
%%%%%%%% AUTHOR CONTRIBUTIONS %%%%%
%%%%%%%%%%%%%%%%%%%%%%%%%%%%%%%%%%%

\begin{contribution}
%%This section gives authors the space to recognize author contributions. The text inside this environment is NOT counted towards the total word quanta. At a minimum, manuscripts are expected to include this text:

N.A. and S.D. contributed equally to this work.

%% But authors are expected to provide more specific details, e.g. 
%%
%%SC was responsible for writing and submitting the manuscript.
%%WWM came up with the initial research concept and edited the manuscript.
%%OTS obtained the funding and edited the manuscript.
%%EBF provided the formal analysis and validation. He also edited the manuscript.
%%GEH Supervised the undergraduates, wrote the software and administers the project github and Zenodo repositories.
%%
%% Authors can use the Contributor Role Taxonomy (CRediT) at
%% for ideas on how write a good statement tailored to their needs.

\end{contribution}

%% in the research for the paper. Each keyword is check against the master 
%% list during copy editing. Individual instruments can be provided in 
% \facilities{HST(STIS), Swift(XRT and UVOT), AAVSO, CTIO:1.3m, CTIO:1.5m, CXO}

%%%%%%%%%%%%%%%%%%%%%%%%%%%%%%%%%%%
%%%%%%%%%%% FACILITIES %%%%%%%%%%%%
%%%%%%%%%%%%%%%%%%%%%%%%%%%%%%%%%%%

%\vspace{5mm}
\facility{STEREO (COR1 and COR2).}

% \software{astropy \citep{2013A&A...558A..33A,2018AJ....156..123A,2022ApJ...935..167A},  
%           Cloudy \citep{2013RMxAA..49..137F}, 
%           Source Extractor \citep{1996A&AS..117..393B}
%           }

%%%%%%%%%%%%%%%%%%%%%%%%%%%%%%%%%%%
%%%%%%%%%%%% SOFTWARE %%%%%%%%%%%%%
%%%%%%%%%%%%%%%%%%%%%%%%%%%%%%%%%%%

\software{The BFF and spectral analysis codes used in this work are freely available on the Zenodo platform \citep{Alzate2025,DiMatteo2025}. The NRGF code is available at \url{https://solarphysics.aber.ac.uk/Data.html}.}


\bibliography{Alzate-DiMatteo-refs}{}
\bibliographystyle{aasjournalv7}

%%%%%%%%%%%%%%%%%%%%%%%%%%%%%%%%%%%
%%%%%%%%%%%% APPENDIX %%%%%%%%%%%%%
%%%%%%%%%%%%%%%%%%%%%%%%%%%%%%%%%%%

\appendix

\section{SPD\_MTM procedure}\label{sec:appendix_spec}
\begin{figure}[h]
    \centering
    \includegraphics[width=0.9\linewidth]{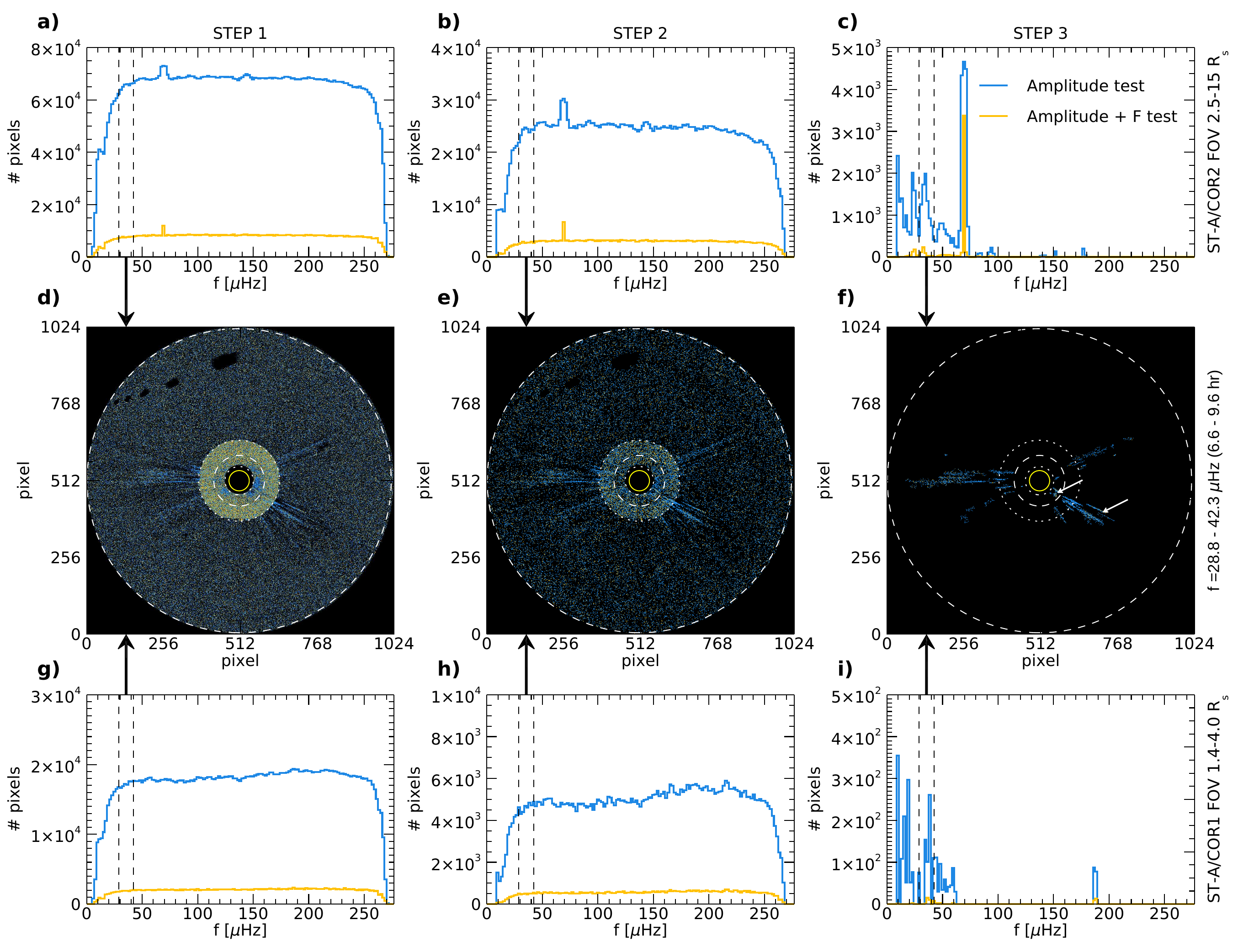}
    \caption{\footnotesize \textbf{Procedure for the reduction of false positive rates.} STEP 1: for each frequency, we collected the location of the pixels at which the amplitude test (blue) and amplitude+F test (orange) is satisfied. STEP 2: for each frequency, we removed isolated pixels and groups with less than three consecutive pixels satisfying the amplitude test; keep the pixels at which the amplitude+F test is also satisfied. STEP 3: for each frequency, we selected groups of at least 25 consecutive pixels or five neighbor groups of three pixels, with maximum one pixel gap, satisfying the amplitude test; keep the pixels at which the amplitude+F test is also satisfied. Panels a--c show the frequency distribution of pixels at each step for COR2 observations. Panels g--i show the same for COR1 observations. Panels d--f show the location of pixels at each step in the $f = 28.8-42.3 \mu Hz$ frequency range (6.6--9.6 hr). The dashed white circle indicates the COR2 FOV, the dotted white circle indicates the COR1 FOV, and the yellow circle indicates the sun.}
    \label{fig:PSD_steps}
\end{figure}

\added{The spectral analysis of brightness time series at a single pixel of coronagraph images was performed using the \emph{SPD\_MTM} routine \citep{DiMatteo2021,DiMatteo2025} based on the aMTM \citep{Thomson1982,Percival1993} combined with additional approaches \citep{Mann1996,Vaughan2005,Viall2015} which make the technique suitable for the analysis of astrophysical and geophysical data series. This method estimates the power spectral density (PSD) of a linearly detrended data series via a weighted average of independent estimates of the PSD based on the Fast Fourier Transform of the data series weighted by $K\leq2NW-1$ orthogonal tapers, specifically the Slepian sequences \citep{Slepian1978}, determined by the user chosen time-halfbandwidth product, $NW$. The choice of the $NW$ parameter determines the smallest resolvable frequency difference of the aMTM method, which is $2NWf_{Ray}$ with $f_{Ray}=1/T$ being the Rayleigh frequency dictated by the length of the time interval, $T$. Panels a and d of Figure \ref{fig:PSD_example} show the brightness time series extracted at two pixels in the COR2 and COR1 FOV (pixels location marked by arrows in Figure \ref{fig:PSD_steps}f). The corresponding estimated PSD are the black lines in panel b and e. Then, via a maximum likelihood criterion (an adaptation of the approach of \citet{Vaughan2005} for red noise spectra), we estimated the continuous PSD background fitting a pan-spectrum \citep{Liu2020} plus constant (PNSc) function:

\begin{equation}
    PSD(f)=cf^{-\beta} \left[1+ \left( \frac{f}{f_{b}}\right)^{a}\right]^{\frac{\beta-\gamma}{a}}+const
\end{equation}

which is a flexible bending power law going from a slope of $\beta$ to $\gamma$ at a frequency break point, $f_{b}$, with sharpness controlled by the parameter \textit{a}. The constant value helps capture the portion of the PSD dominated by shot noise. Then, periodicities in the time series are identified at frequencies at which the ratio of the PSD with respect to the PSD background (solid red lines), which follow a gamma distribution with $\approx2K$ degrees of freedom \citep{DiMatteo2021}, shows enhancements above a 90\% confidence threshold (dashed red lines). While this ``amplitude test'' is sufficient for the identification of quasi-periodic variations, the aMTM provides an additional independent test, called F test, which examines for the presence of monochromatic and phase coherent sinusoidal signals. The F test is based on a random variable (F values, panel c and f), following a Fisher distribution of $2K$ degrees of freedom. While this test should not be used by itself because it is prone to identify false positives \citep{Protassov2002}, when combined with the amplitude test, it provides reliable results. The level of false positives for each test and their combination have been validated by numerical simulations inside a ``safe'' frequency range, $[2NWf_{Ray},f_{Ny} - 2NWf_{Ray}]$ \citep{DiMatteo2021}, where the Nyquist frequency, $f_{Ny}=1/(2\Delta t)$, is determined by the time series sampling rate, $\Delta t$. By imposing a confidence threshold $C$, we expect a false positive rate of $1-C$ for each test individually and $\approx (1-C)/(4NW)$ when both amplitude and F tests are satisfied \citep{DiMatteo2021}.

\begin{figure}[t]
    \centering
    \includegraphics[width=0.9\linewidth]{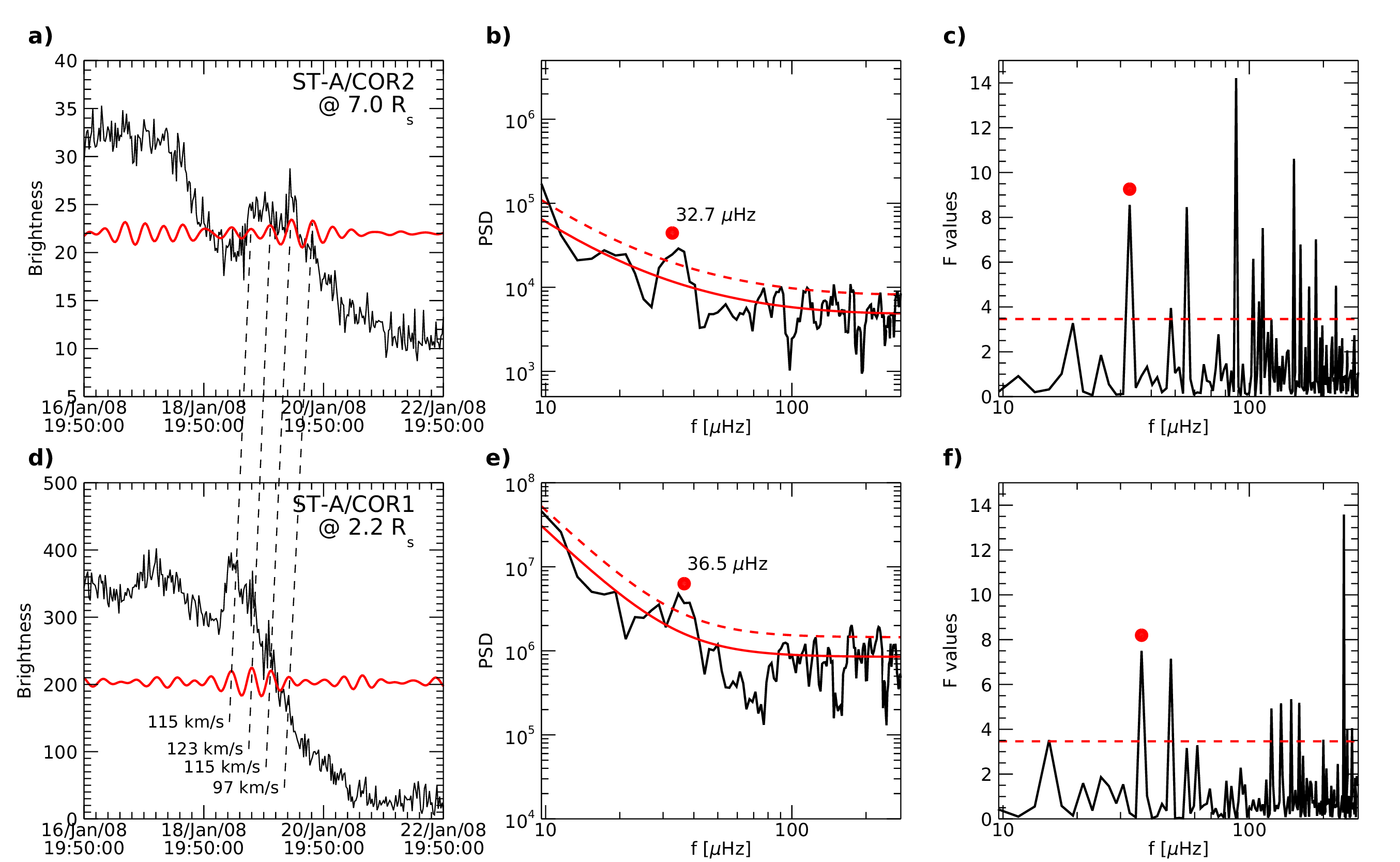}
    \caption{\footnotesize \textbf{Spectral analysis results of brightness time series from two pixels indicated by arrows in Figure \ref{fig:PSD_steps}f.} For COR2 observations: panel a shows the brighness time series compared to the data filtered at the identified frequency (red); panel b shows the PSD with the identified PSD background (solid red) and the 90\% confidence threshold (dashed red), panel c shows the F values with the 90\% confidence threshold (dashed red). The red circle marks the frequency of an identified periodic signal satisfying both the amplitude and F test. Panels d--e show the same results but for COR1 observations. We associate similar crests in the brightness variations (black dashed lines) and provide rough estimates of their propagation speed.}
    \label{fig:PSD_example}
\end{figure}

\section{Refinements for coronagraph observations}\label{sec:appendix_refine}

We test our approach on a 6 day interval of STEREO-A/COR2 and COR1 observations, respectively in the FOV 2.5-15 \Rs\ and 1.4-4.0 \Rs, from 2008/01/16 19:50 to 2008/01/22 19:50 UT with time step of $\Delta t=$30 min. In Figure \ref{fig:PSD_steps}, we show the frequency distribution of the detected signals via the amplitude test (blue) and amplitude+F test (orange) for COR2 (panel a) and COR1 (panel g) observations. The location of such identification in the frequency range $f=28.8-42.3 \mu Hz$ is displayed as a composite with the same colors in panel d. The dashed white circles delimit the COR2 FOV, while the dotted white circles delimit the COR1 FOV. The sun is represented by the solid yellow line. Based on the parameter choice in our investigation ($NW=2.5$ and $C=0.90$) and for an image of dimension of $1024\times1024$ (COR2 images) or $512\times512$ (COR1 images) pixels, the expected level of false positive is $\approx 10^5$ or $\approx2.6\times10^4$ for the amplitude test and $\approx 10^4$ or $\approx2.6\times10^3$ when the two tests are combined. The distribution of pixels remains relatively flat in the ``safe'' frequency range, $[9.6,268.1] \mu Hz$, with median values of $6.8\times10^4$ (COR2) and $1.8\times10^4$ (COR1) pixels for the amplitude test and $8.2\times10^3$ (COR2) and $2.1\times10^3$ (COR1) pixels for the amplitude+F test. These values are close to our expectations, given that not all pixels in the image are actually analyzed (outside the chosen FOV) and some areas affected by artifacts result in large dark patches within the FOV (panel d). A peak at $f=69.4 \mu Hz$ ($\approx 4 hr$) stands out for COR2 observations and corresponds to an already reported spurious periodicity limited only to regions close to the inner FOV \citep{Viall2015,Alzate2024}. Due to the high level of false positives, peaks in the frequency distribution are relatively weak. By looking at the location of such identification in the $f=28.8-42.3 \mu Hz$ frequency range (panel d), there are evident clusters of pixels distributed along fine-radial structures. We capitalize on this property to reduce the level of false positives. After collecting the groups of consecutive pixels at which the amplitude test (blue dots) is satisfied (``true'' pixels) at each frequency (STEP 1), we remove the ones with less than three ``true'' pixels (STEP 2) and obtained the distributions in panels b, e, and h. Lastly (STEP 3), for each frequency, we surround the remaining ``true'' pixels with temporary positive pixels and use 8-neighborhood pixel connectivity to select groups of at least 74 pixels. For this condition to be satisfied, we need at least five groups of three consecutive ``true'' pixels separated by no more than one pixel from each other. In addition, we also consider single groups of at least 25 ``true'' pixels. This approach is very selective and retains only the more reliable features as shown in panel f. At each STEP, within the features identified by the amplitude test, we also record the pixels at which the amplitude+F test is also satisfied (orange dots). The final distributions, shown in panels c and i, reveal the presence of preferred frequency bands. The pixel locations in panel f reveal fine quasi-radial features in the COR2 FOV and possibly streamer boundaries in the COR1 FOV.

As an additional check of our results, we compare the brightness time series from two pixels at which we detected a similar periodicity (white arrows in panel f). The identified periodicity, highlighted by bandpass filtering the time series at the identified frequency (red lines in panels a and d in Figure \ref{fig:PSD_example}), shows the same structures at the two locations. Given the height of the selected pixels and the timestamps of the oscillation crests, matched by dashed black lines, we estimated average propagation speeds ranging from $\approx$97 to $\approx$123 km/s for these features. Our rough estimates are comparable to speeds associated with propagating features spanning the low-, middle-, and high-corona \citep[][and references therein]{Alzate2024} further supporting the reliability of our analysis.
}

\end{document}